\documentclass[12pt,preprint]{aastex}
\usepackage{graphics}

\def\gtorder{\mathrel{\raise.3ex\hbox{$>$}\mkern-14mu
             \lower0.6ex\hbox{$\sim$}}}
\def\ltorder{\mathrel{\raise.3ex\hbox{$<$}\mkern-14mu
             \lower0.6ex\hbox{$\sim$}}}

\def\Msun{\>{\rm M_{\odot}}}

\shorttitle{Stability and Mass Loss}
\shortauthors{Debes \& Sigursson}

\begin{document}
\title{Are There Unstable Planetary Systems Around White Dwarfs?}
\author{John H. Debes \& Steinn Sigurdsson}
\affil{Pennsylvania State University}
\affil{Department of Astronomy \& Astrophysics, University Park, PA 16803, USA}

\begin{abstract}

The presence of planets around solar-type stars suggests that many white dwarfs 
should have relic planetary systems.  While planets closer than $\sim$ 5~AU will 
most likely not survive the post-main sequence lifetime of its parent star, any 
planet with semimajor axis $>$ 5~AU will survive, and its semimajor axis will 
increase as the central 
star loses mass.  Since the stability of adjacent orbits to mutual planet-planet 
perturbations 
depends on the ratio of the planet mass to the central star's mass, some planets 
in previously stable orbits around a star undergoing mass loss will become 
unstable.  
We show that when mass loss is slow, systems of two planets that are marginally 
stable can become unstable to close encounters, while for three planets the 
timescale for close approaches decreases significantly with increasing mass ratio.  
These processes could explain the presence of anomalous IR excesses around 
white dwarfs that cannot be explained by close companions, such as G29-38, and
may also be an important factor in explaining the existence of DAZ white 
dwarfs.
The onset of instability through changing mass-ratios will also be 
a significant effect for planetary embryos gaining mass in protoplanetary disks.  
\end{abstract}   


\section{Introduction}
The discovery of $\sim$ 80 planets, and counting, around solar type stars 
suggests that 
successful planet formation is quite common.  The wealth of systems so unlike 
the Solar
 System leads one to conclude that many aspects of planetary system 
formation and dynamical evolution have yet to be fully explored.  
One particularly interesting area is the long term evolution of planetary 
systems specifically in the presence of post-main-sequence evolution of the central star. Observations of 
planets around post-main-sequence stars may provide additional information about the 
formation and evolution of planetary systems around main-sequence stars, and can inform us 
about the long term future of the solar system.

While planets at distances similar to the outer planets in the solar system will 
persist through post-main-sequence evolution \citep{1998Icar..134..303D}, it is 
unlikely that close Jovian companions to such stars survive.  As the star 
evolves 
it expands, engulfing anything up to 
$\sim$ 1~AU \citep{1993ApJ...418..457S,1999MNRAS.304..925S,1999MNRAS.308.1133S}.  
Outwards of an AU, up to $\sim$ 5~AU any planet's orbit will decay through 
tidal transfer of angular momentum and become consumed within the envelope 
of the star \citep{1996ApJ...470.1187R}.  Anything with less mass than a brown 
dwarf will not survive in the stellar envelope 
\citep{1984MNRAS.208..763L,1984MNRAS.210..189S}.

Planets may still be observed in close orbits around white dwarfs if their 
orbits are significantly changed by some process that occurs after the AGB 
phase.  
If the planets become unstable to close approaches with each other, their 
interaction would result in a planet close to the central star, similar to 
scenarios proposed for the formation of close Jovian planets around main 
sequence 
stars \citep{1996Sci...274..954R,1996Natur.384..619W,1997ApJ...477..781L}.  
The results of close encounters between two Jovian planets have been studied in 
detail, with three possible outcomes: the two planets collide leaving a large 
planet,
 one planet is ejected, or both planets remain in a new stable configuration 
\citep*{havford}.  
For planets starting out with semi-major axes $>$ 5~AU, $\sim$8\% of unstable 
pairs will collide, the remaining $\sim$92\% will not. 
Of the systems which avoid collision, roughly 40\% will settle into a 
configuration with a planet in a 
significantly closer and more eccentric orbit than in the initial system.
Thus the onset of instability after post-main-sequence mass loss may create
white dwarf systems with planets at obital radii scoured clear of the original 
inner planets
during the star's giant phase.

\citet{1997Icar..125....1D}, simulating the Uranian satellite system, found 
that Hill stable systems can become unstable with 
an increasing mass ratio for satellites orbiting a central massive object.
This important work led to one of the few systematic studies of the 
post-main-sequence evolution of planetary systems dynamically similar 
to the solar system \citep{1998Icar..134..303D}.  In that paper, Duncan \& 
Lissauer
found that the time to unstable close approaches for the planets followed 
a power-law relationship with the ratio of planetary mass to stellar mass, as
an increasing fraction of mass was lost from the central star.
At the level of individual planetary orbits, resonances may also play an 
important role 
in an adiabatically changing 
system, enhancing stability or creating instability.  
In this paper we generalize the specific results of \citet{1998Icar..134..303D} 
to a range of system parameters applicable to a wider range of situations, such 
as those  
like the multi-planet extrasolar systems recently discovered.

Most likely, extrasolar planetary systems also possess Oort cloud analogues as a 
direct result of planet formation 
\citep{1950BAN....11...91O,1999SSRv...90..301W}.    
For comets in the outer Oort cloud orbital time scales are comparable to the
more rapid late stages of post-main-sequence evolution, and the mass loss of a 
star is not adiabatic in the 
context of AGB evolution.  If the mass loss is fairly symmetric, many 
of these comets survive the evolution and can later provide a steady flow of 
comets which  
impact the white dwarf directly, or break up due to tidal strain and populate 
the inner system
with dust,
causing photospheric metal contamination 
\citep{stern,1986ApJ...302..462A,1998ApJ...501..357P}.  
However, if the planet systems become unstable to close approaches after the AGB 
phase, the entire 
system becomes dynamically young and new collisions and encounters can occur 
between surviving comets and planets. Many scenarios lead to a period
of enhanced ``late bombardment'' as cometary orbits are perturbed and the
flow of comets into the inner system is enhanced.  In this paper we will 
investigate whether 
this mechanism can explain the observed IR excess around the white dwarf G29-38, 
attributed to a disk of dust whose extent is comparable to the tidal radius of 
the white
dwarf \citep{graham,1987Natur.330..138Z}.  \citet{graham} estimated that 
approximately 10$^{15}$ g/yr of 
metal rich material would rain upon the white dwarf, if the infrared emission
is due to a dust cloud at about half solar radius, requiring a steady and
high rate of replenishment of the putative dust.  Our model may resolve the 
problem of those
DAZ white dwarfs which cannot be easily explained by
either isolated cometary impacts or ISM accretion \citep{1998ApJ...505L.143Z}.

We will show that the mass lost from a central star is sufficient to destabilize 
systems of two or more planets in previously stable orbits and cause them 
to suffer close approaches, producing several observable signatures. 
In section 2 we will develop the stability of 
planetary systems against close encounters in the presence of adiabatic mass 
loss, 
describe our numerical methods for testing our analytical estimates in section 
3, 
present our results and discuss relevent observational signatures in section 4, 
and 
discuss the implications of these findings in section 5.

\section{Stability for Planetary Systems Under Mass Evolution}
\subsection{Two Planet Systems}

The stability of two planets against close approaches depends primarily on the 
masses of the planets relative to the central star and the separation of the two 
orbits.  
This separation is measured as $\Delta=(a_{2}-a_{1})/a_{1}$ where $a_1$ and 
$a_2$ are 
the inner and outer semi-major axes respectively.  A critical Hill separation, 
$\Delta_c$ 
is then the minimum separation between two planets which ensures a lack of close 
approaches 
over all time \citep{hill}. A full treatment of the Hill stability of two 
planets in the 
case of static masses can be found in \citet{1993Icar..106..247G}.  
Several approximations can be made that simplify the full treatment, 
such as equal planetary masses and small eccentricities.  The criterion is then 
given by:
\begin{equation}
\label{deleqn}
\Delta_c \simeq \sqrt{\frac{8}{3}\left(e^2_1+e^2_2\right)+9\mu^{2/3}}
\end{equation}
where $\mu$ is the ratio of the planets' mass to the central star, $e_1$, $e_2$ 
are 
the eccentricities of planet 1 and 2, and $\Delta_c$ is in units of the inner 
planet's semi-major axis $a_1$.  

If either the mass of the planets or the mass of the star changes, the critical 
Hill 
radius will change as well.  An increase in planet mass or a decrease in stellar 
mass 
will cause $\mu$ to become larger, increasing the width of the zone in which 
orbits are 
unstable to close approaches.  During post main-sequence mass loss, the orbits 
of planets 
will widen as the central star loses mass.  As long as this process is 
adiabatic, 
the planets will simply conserve their angular momentum and widen their orbits 
proportionally to the mass lost: $a_{new}=a_{old}(M_{i}/M_{f})$.  
However, since the orbits widen together by the same factor, $\Delta$ remains 
the same.  
Thus, while the critical separation at which the two planets will become 
unstable widens, 
their relative separation remains unchanged.  Orbits that are initially 
marginally stable,
or close to being unstable,
will become unstable to close planet--planet approaches as a consequence of the 
mass loss
from the central star.  
In the case of planetary mass accretion in a protoplanetary disk, 
the orbits of the two planets will remain the same while $\Delta_c$ increases, 
creating the same effect as if the star were losing mass.

The opposite case of stellar mass accretion or planetary mass loss works to make 
previously unstable regions more stable.  However, since close approaches 
generally 
happen within a few tens of orbits, objects likely would be cleared out of an 
unstable 
region quicker than the region could shrink.

\subsection{Multiple Planet Systems}

We expect that multiple planet systems should be common, e.g. the solar system, 
PSR 1257+12, and $\upsilon$ Andromedae 
\citep{1994Sci...264..538W,1999ApJ...526..916B}, and thus it is useful to 
develop an idea of how these systems remain stable.  \citet{1996Icar..119..261C} 
found a relation between the separation of a system of planets and the time it 
would 
take for the system to suffer a close encounter:
\begin{equation}
\label{eqn2}
 \log{t} = b\delta+c
\end{equation}
where $b$ and $c$ are constants derived through numerical simulations.  
The symbol $\delta$ is related to $\Delta$ but is defined in a slightly 
different way.  
Here, $\delta$ is the separation between two planets ($a_{i+1}-a_{i}$) in units 
of mutual 
Hill radii (R$_i$) defined as such:
\begin{equation}
\label{rh}
R_i=\left( \frac{1}{3}\mu \right)^{1/3} \frac{\left( a_{i+1}+a_{i} \right)}{2}
\end{equation}
where $i$ can be from 1 to $N_{planets}-1$ and we assume the planets have equal 
masses and initially circular orbits.  If the parameter $\delta$ is the same for 
each pair of adjacent planets the separations in units of AU will be different.  
For example, if we took three Jovian mass planets ($\mu\sim10^{-3}$) with 
$\delta=6.5$ 
and the innermost Jovian at 5.2~AU from the central star, the next two planets 
would be 
at 9.4 and 16.7~AU. Compare to actual orbital radii of 9.6 and 19.1 AU for 
Saturn and Uranus 
respectively, in our presumably stable for several billion years solar system.
We add the obvious caveat that Saturn and Uranus are significantly less massive than Jupiter
and have correspondingly weaker mutual interactions.

Adiabatic mass evolution will have the effect of shortening the time it takes 
for orbits to suffer close approaches.  The knowledge of this has long been used to 
speed up numerical calculations \citep[][and references 
therein]{1997Icar..125....1D}.  However, this fact 
also leads to the hypothesis that planetary systems on the edge of stability for 
10$^{10}$ yr 
will be affected by mass loss.  In general the new time to close aproaches for 
an 
initial $\delta$ with a change in mass is given by:
\begin{equation}
\log{\frac{t_f}{t_i}} = (b^\prime-b)\delta+(c^\prime-c).
\end{equation}
We would expect $c$ to have little or no change with a change of mass since it 
represents the timescale for two planets at $\delta\sim$0 to suffer a close 
approach.  
Mass loss will increase the mutual Hill radii of the planets which in turn will 
change $b$ 
to a new value we will define as $b^\prime$:
\begin{equation}
\label{bprime}
b^\prime=\left(\frac{\mu_f}{\mu_i}\right)^{\frac{1}{3}}b
\end{equation}
where $\mu_f$ and $\mu_i$ are the final and initial mass ratios respectively.  
Such behavior suggests that bodies that are stable over the lifetime of a 
planetary system 
will become unstable over a timescale several orders of magnitude smaller than 
their
original timescale for instability, when the central 
star becomes a white dwarf, assuming the relation of Equation \ref{eqn2} holds 
for large $\delta$.  
It has been found that for the case of three planets with $\mu$=10$^{-7}$, 
the parameters $b\simeq1.176$ and $c\simeq-1.663$ \citep{1996Icar..119..261C}.  
If the three planets are each separated from their neighbor by $\delta$=6, they 
will 
experience close encounters after $\sim$10$^5$ orbits of the inner planet.  
For comparison, three planets with the same mass ratio and separated by a 
$\delta=8$ 
will experience close encounters after $6 \times 10^7$~yr.  Assuming the central 
star 
loses half of its mass, the timescale to close encounters will shorten by an 
order of 
magnitude for the first case and two orders of magnitude for the second.  

\subsection{When is Mass Evolution Adiabatic?}
The question of whether mass evolution is adiabatic needs to be addressed. 
In the case of mass loss for solar mass stars, roughly half the central star's 
mass will be lost on the order of $10^{8}$ yr.  A majority of the mass is lost 
at 
the tip of the AGB branch during a period of $\sim$10$^6$ yr.  Even the 
quickest 
rate of mass loss is much longer than one orbital period of a planet inwards of 
100 AU, 
the general region where planets are believed to have formed.  Stars heavier 
than a 
solar mass probably have superwinds which will cause significant mass loss on 
the order 
of a few hundred or thousand years 
\citep{1993ApJ...413..641V,1999A&A...349..898S}.  
Whether this is important or not will be the subject of further study.  
Objects very far away from the central star, such as Oort cloud object 
analogues, 
have orbital timescales comparable to the mass loss timescale and will also not 
follow the adiabatic case.  It should be noted that for Kuiper and Oort cloud 
distances that the
mass loss by the star would become adiabatic if the asymptotic wind velocity 
were orders
of magnitude smaller than the escape velocity at the surface of the star since 
the 
crossing time of the wind would then be larger than the orbital timescale of the 
comets.  

Mass gain of stars and planets is much slower than the orbital timescale of a 
planet.  
Accretion rates for protostars are on the order 
$10^{-6}$ M$_{\odot}$ yr$^{-1}$ \citep{1987ARA&A..25...23S}.  
The formation of giant planets through runaway gas accretion takes 
$\sim$10$^{7}$ yr, 
the rough lifetime of gaseous protoplanetary disks \citep{1996Icar..124...62P}.  
If some giant planets are formed more quickly by more efficient runaway 
accretion, 
gravitational collapse \citep{2000ApJ...536L.101B}, or seeding through the 
formation 
of other planets \citep{1999Natur.402..633A}, they would not be described by the 
adiabatic case. 

\section{Numerical Methods}  
In order to test the hypothesis that adiabatic mass evolution should change the 
stability of planetary systems, we ran several numerical simulations of two 
planet 
and multi-planet systems in circular orbits around a central star losing mass.  
The equations of motion were integrated using a Bulirsch-Stoer routine 
\citep{bs,numrec}.  
Since the case of mass loss of the central star and mass gain of the planet is 
the same, 
mass loss can be modeled in two ways.   Either the star's mass can be decreased, 
or the 
planets' masses can be increased.  If the planets' masses are increased the time 
coordinate 
must be scaled to reflect the fact that the orbits are widening.  To keep our 
investigations 
scalable, we chose the units of time to be orbits of the inner planet.   We 
chose to 
increase the mass of the planets over a period of 1000 orbits.  In the absence 
of 
mass evolution, energy and angular momentum were conserved to better than 1 part 
in 10$^6$ for 10$^5$ orbits.  Since changng mass makes this a 
non-conservative system, 
energy and angular momentum could not be used as a test of accuracy.  However, 
since the simulations were integrated until a close approach and then 
terminated, 
any error is similar to the case of no mass evolution.  Several simulations 
without 
mass evolution were run with stable results.  A close approach was defined by an 
encounter separated by a radius of $<$ 2$\mu^{2/5}$ \citep[][and references 
therein]{1993Icar..106..247G}.  
This radius was chosen because at separations smaller than this the 
planet-planet 
system is dominant and the star becomes a perturbation.  
Other authors have chosen different criteria \citep{1996Icar..119..261C}, but 
the 
results are insensitive to the exact choice.  

In the two planet case, we started simulations at the critical separation 
predicted 
by Equation \ref{deleqn} assuming no mass loss, and increased the separation 
between 
the two planets at regular intervals in $\Delta$.  We integrated the equations 
of 
motion until a close approach or for 10$^5$ orbits.  We increased $\Delta$ until 
it was 25\% greater than what would be predicted in the presence of mass loss.  
These simulations were run an order of magnitude longer than 
\citet{1993Icar..106..247G}, 
and in the no mass loss case were consistent with what they found.  The two 
planets were 
initially started with true anomalies separated by 180$^\circ$.  Our separations 
are then 
lower limits for the critical separation and thus truly reflect the minimum 
possible separation 
between orbits that remain stable.   For multiple planets, $\delta$ was started 
at 2.2 and 
raised until several consecutive separations did not experience close encounters 
for 10$^7$ orbits.  
Here random phases in the orbits where chosen with the restriction that adjacent 
orbits were 
separated by at least 40$^\circ$.  Three separate runs with different random 
initial phases 
were performed to improve the statistics for each mass, as there is significant 
scatter 
in the actual time to a close approach for each separation.

\section{Results}
\subsection{Two Planets}
\begin{figure}
\plotone{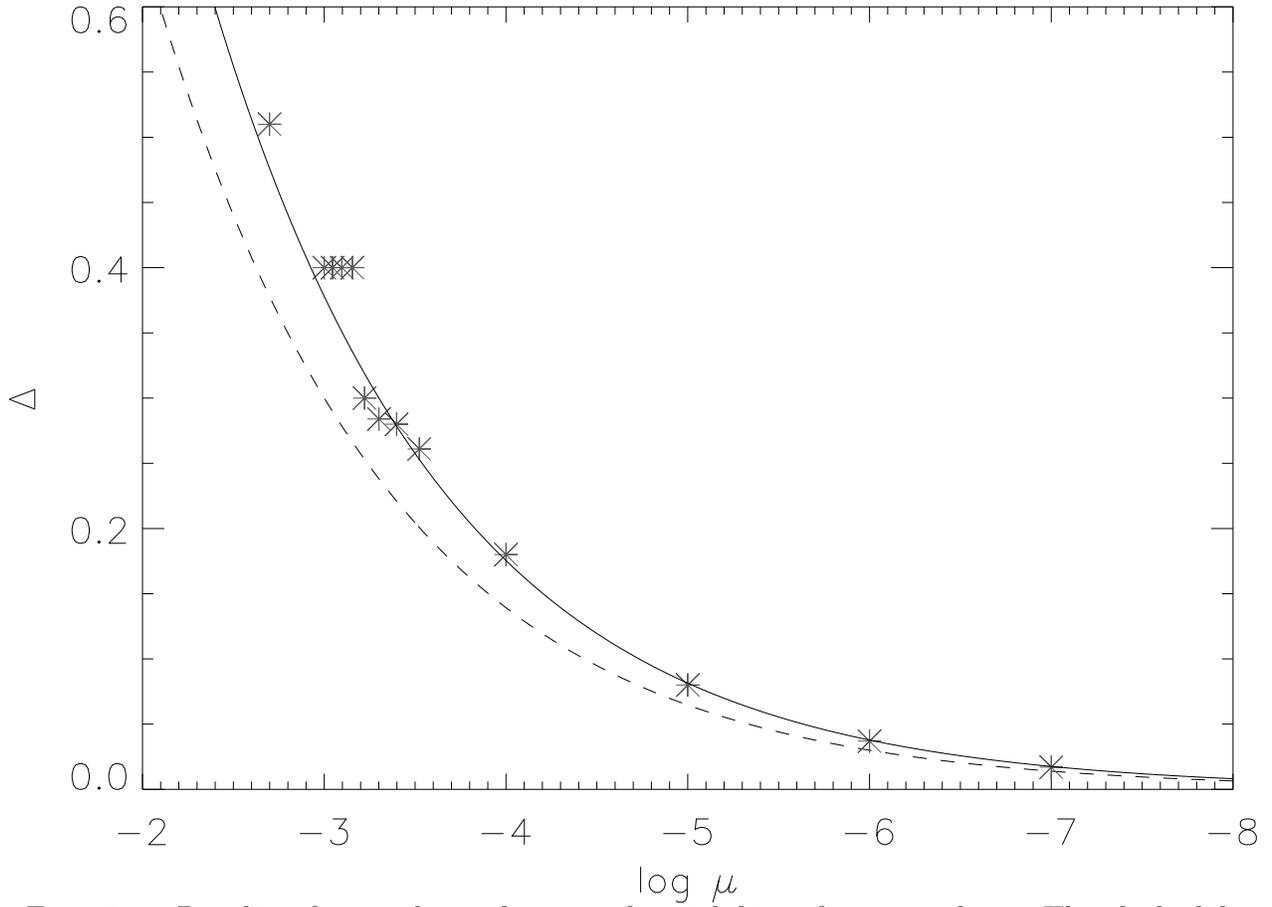}
\caption{\label{logstab} Results of several simulations of instability after 
mass loss.  The dashed line corresponds to the original stability criterion of 
Equation \ref{deleqn}.  The solid line corresponds to the criterion with 
$\mu=2\mu_i$}
\end{figure}

We looked at a wide range of planetary masses for a solar mass star, from a sub 
terrestrial-sized planet ($\mu=10^{-7}$) to a Jovian planet ($\mu=10^{-3}$).  
Figure \ref{logstab} shows the border for onset of instability in two planet
systems after mass loss. The dashed line represents the 
initial critical Hill radius for no mass loss.  The solid line which goes 
through the 
points is the critical Hill radius for $\mu$ equal to twice that of the initial 
system, 
corresponding to the planets doubling in mass or the central star losing half of 
its mass.  
Several of the higher mass points are greater than that predicted by the solid 
curve, 
an indication of higher order $\mu$ terms becoming important.  It should be 
noted that 
these results are general to any combination of planet and stellar mass that 
have these ratios.

\begin{figure}
\plotone{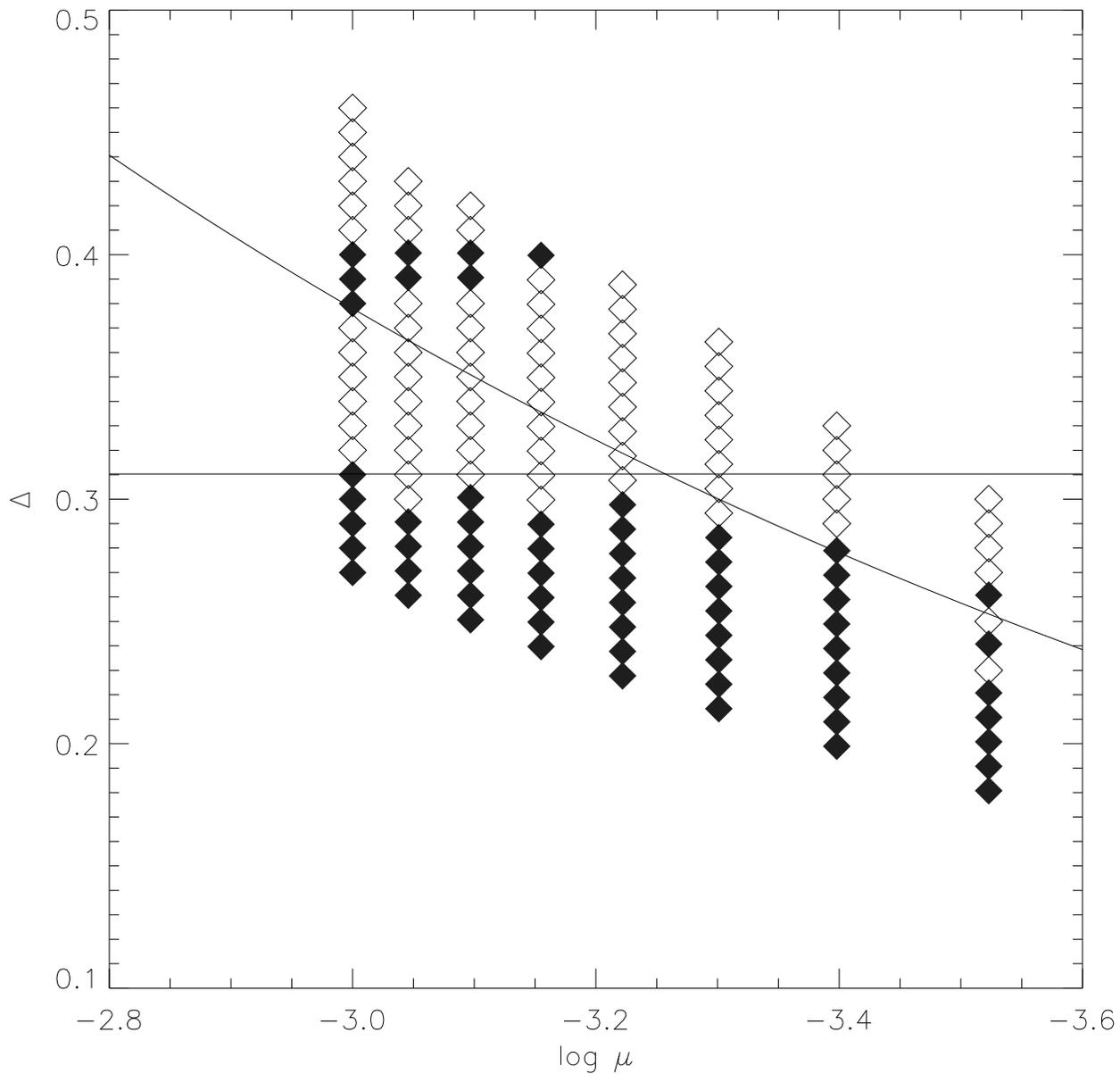}
\caption{\label{island} Close-up of Figure \ref{logstab} in the range of 
10$^{-3.5}$-10$^{-3}$.  Open diamonds represent simulations that did not suffer 
any close approaches over 10$^{5}$ orbits.  Closed diamonds represent 
simulations that did suffer a close approach, while the solid curve represents 
the predicted $\Delta_c$ with mass loss.  The horizontal solid line shows the 
relative separation that corresponds to the 3:2 resonance.  A region of 
stability where instability is expected surrounds this resonance.}
\end{figure}

\begin{figure}
\plotone{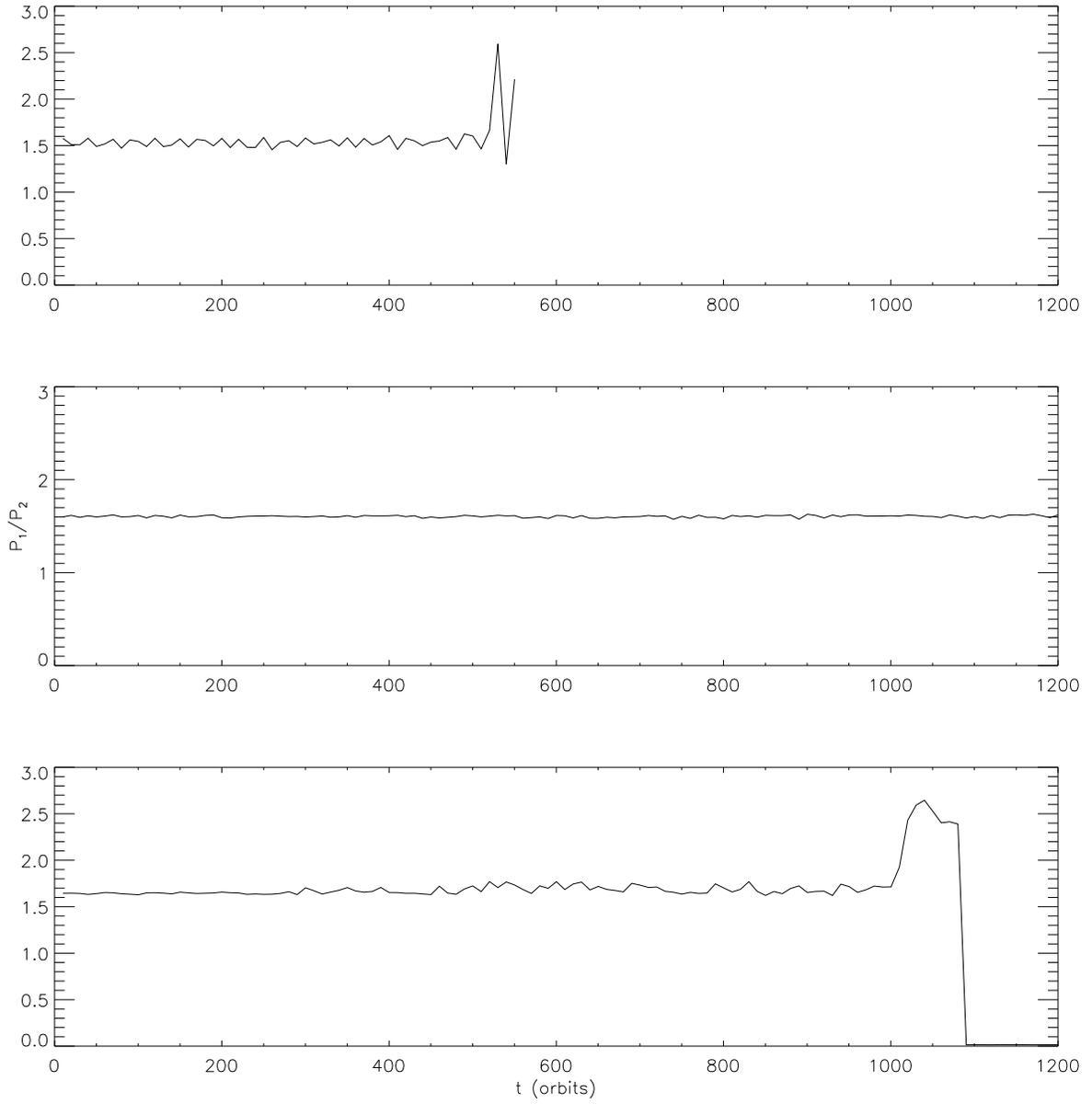}
\caption{\label{resonance} Orbits around the 3:2 resonance for two planet 
stability.  The top and bottom lines correspond to orbits that end in a close 
approach while the middle line shows an orbit that was stable over the length of 
the simulations.}
\end{figure}

In a few cases, separations predicted to become unstable after mass loss by the 
Hill 
criterion were stable for the length of our simulations.  Particularly in the 
$\mu$=10$^{-3}$ case, 
there was a large region in which the two planets suffered no close encounters 
(See Figure \ref{island}).  
These orbits corresponded to a range of $\Delta$ from .32 to .37, 
which were predicted to be unstable under mass evolution from the simple scaling 
of the equation 
for $\Delta$.  It is interesting to note that all of these orbits are close to 
the 3:2 
resonance (See Fig. \ref{resonance}).  For the same reason that the Hill radius 
will not change, 
these orbits will retain the ratio of their periods.  The reason for the 
stability 
around the 3:2 resonance may be due to those separations being near but not in a 
region 
of resonance overlap \citep{1980AJ.....85.1122W,2001Natur.410..773M}, clearly 
this conjecture needs to be confirmed but that goes 
beyond the scope of this particular paper.

\subsection{Multiple Planets}
\begin{figure}
\plotone{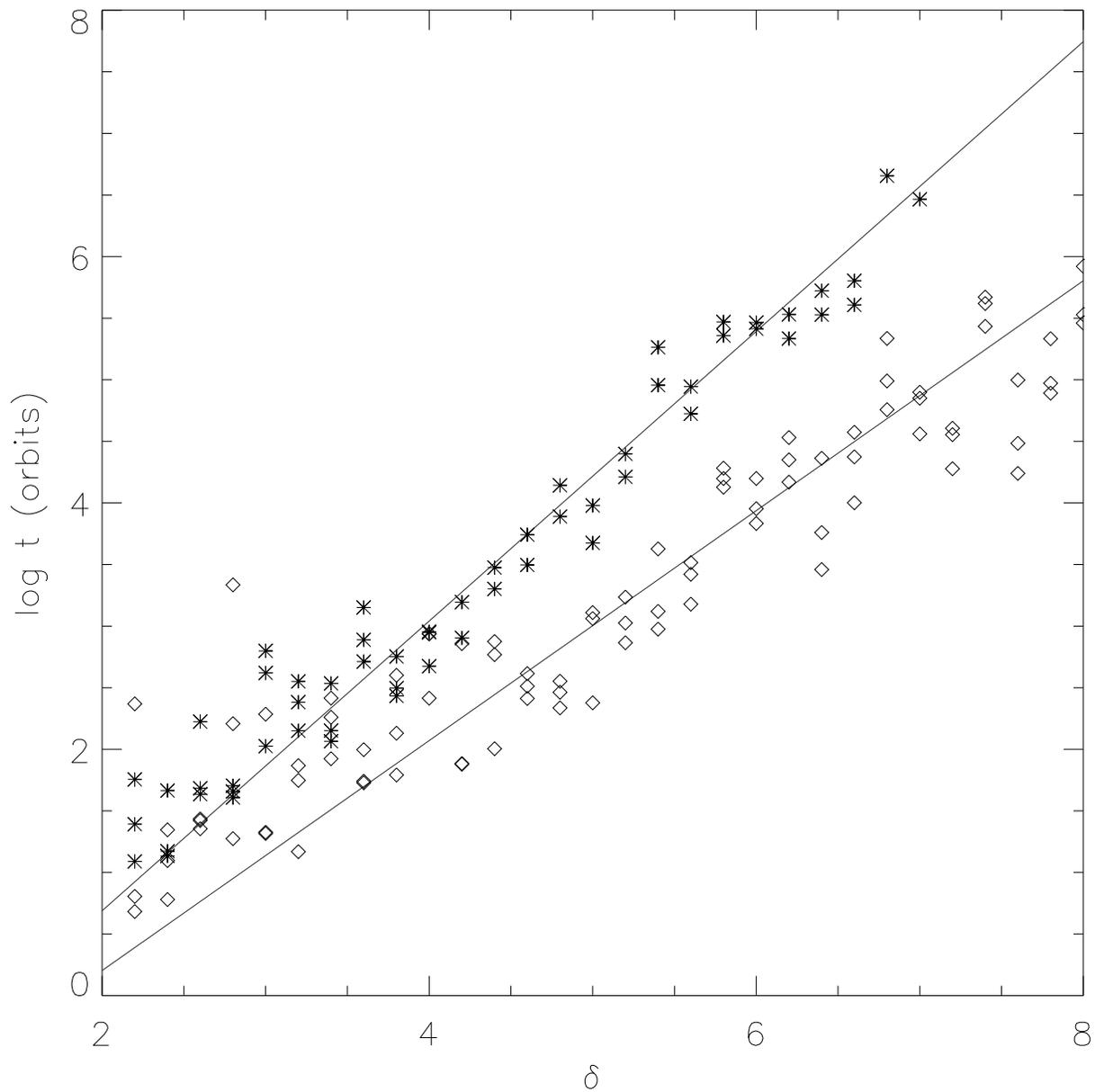}
\caption{\label{multi1} Comparison of the timescale to the first close approach 
for a system of three $\mu$=10$^{-7}$ planets with and without mass loss, where 
stars represent static masses and open diamonds represent the presence of mass 
loss.  The top line is given by least-squares fitting a line of slope $b$ and 
intercept $c$ for no mass loss.  The bottom line is given by Equation 
\ref{bprime}.}
\end{figure}

\begin{figure}
\plotone{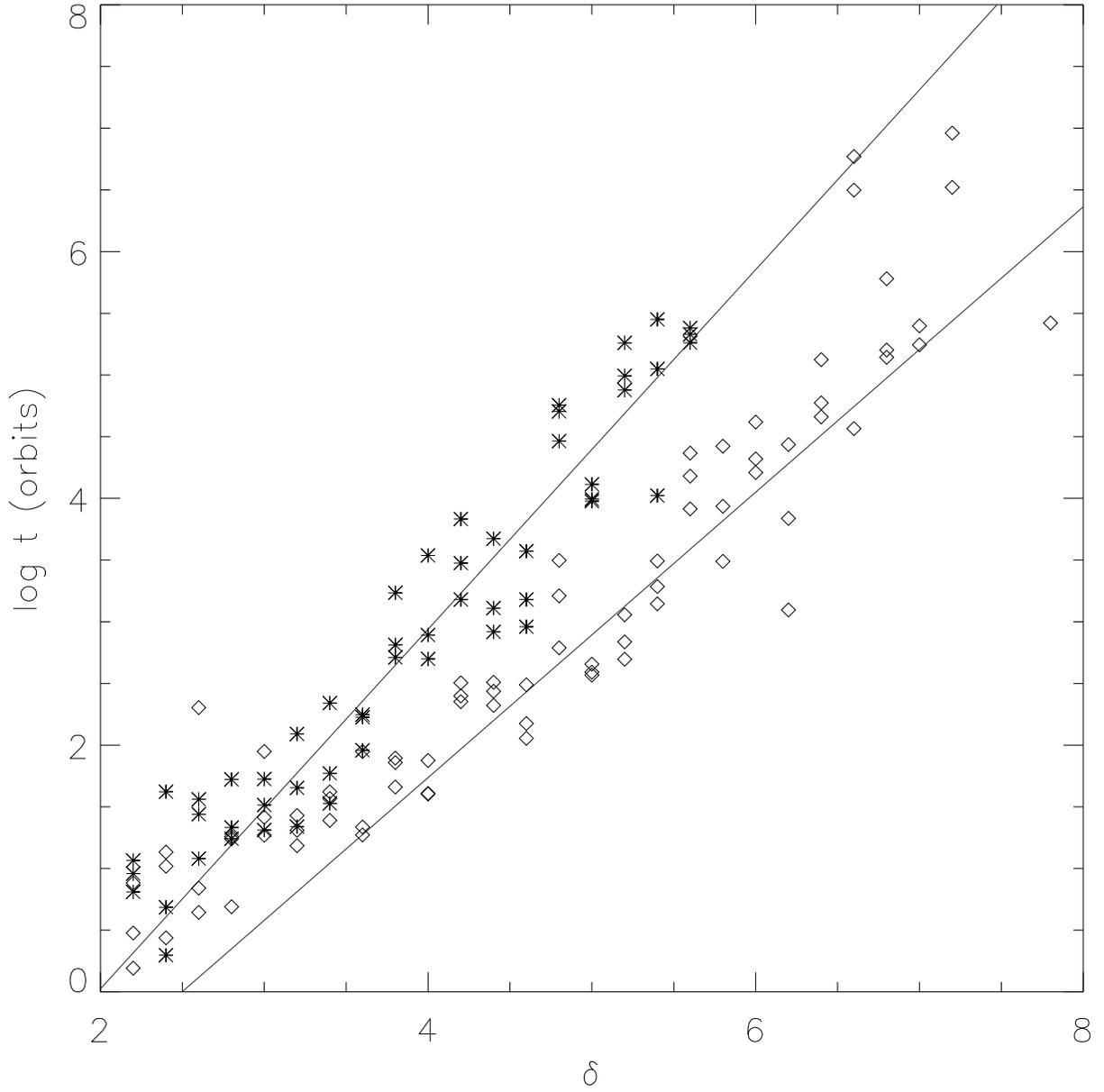}
\caption{\label{multi2}The $\mu=10^{-5}$ case, symbols the same as in Figure 
\ref{multi1}.  The slope and intercept of the top line was derived by fitting 
the numerical simulations without mass loss.  The slope of the bottom line is 
the predicted change due to mass loss.}
\end{figure}

\begin{figure}
\plotone{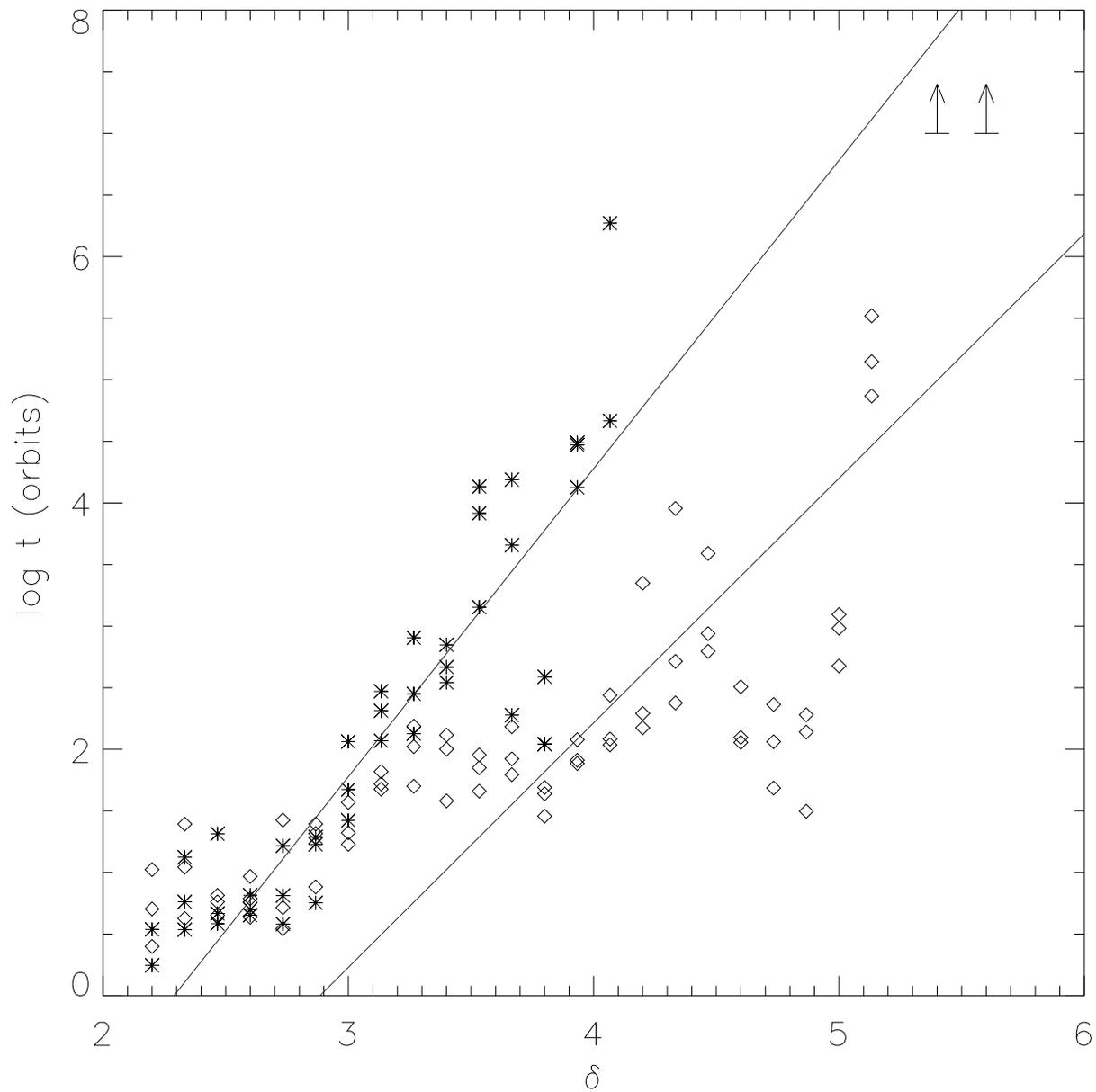}
\caption{\label{multi3} The $\mu$=10$^{-3}$ case, symbols the same as in Figure 
\ref{multi1}. The arrows indicate separations at which our simulations remained 
stable for 10$^7$ orbits.  The slope and intercept of the top line was derived 
by fitting the numerical simulations without mass loss.  The presence of strong 
resonances is particularly noticeable as enhanced stability around $\delta$=5.2 
for the mass loss case, which corresponds to the 2:1 resonance.}
\end{figure}

Figures \ref{multi1}-\ref{multi3} show the results for three different runs, 
looking 
at three planet systems in circular orbits.  We looked at the mass 
ratios $\mu$=10$^{-7}$, 10$^{-5}$, and 10$^{-3}$.  The results are compared to 
simulations without mass loss, and the difference between the two is quite 
noticeable 
for the whole range of mass.  It is important to note that for separations whose 
time to close 
approach is comparable to the mass loss timescale show little change in behavior 
between 
the two cases.  This is because the change in the time to close approach is 
smaller than 
the scatter in the simulations.  Least squares fitting of the static and mass 
loss cases 
were performed to get the coefficients $b$, $c$, and $b^\prime$.  To test our 
assumption 
of $c$ not changing under mass evolution, we also measured $c^\prime$, the 
intercept for 
the mass loss case.  Planets with initial separations in $\delta$ that were 
less than $2\sqrt{3}$, the two planet stability criterion in units of $R_i$, 
were discarded.  
For the mass loss case, points where the timescale of close approaches were 
comparable to the mass loss timescale were also discarded.  Once the 
coefficients were 
determined they were compared to what was predicted from Equation \ref{eqn2}.  
Similarly, $b$ and $c$ from the $\mu$=10$^{-7}$ case without mass loss were 
compared 
with the results of \citet{1996Icar..119..261C}.  Table \ref{tab1} shows that 
within 
the uncertainties, $c$ indeed does not change with mass evolution and the slopes 
are 
consistent with predictions.  Additionally, our results for the static case 
with $\mu$=10$^{-7}$ are consistent with \citet{1996Icar..119..261C}'s values 
for $b$ and $c$.  

As mass increases, the presence of strong resonances becomes more important.  
This is 
due to our choice of equal separations and equal masses, many of these 
resonances would 
disappear with small variations in mass, eccentricity, and inclination 
\citep{1996Icar..119..261C}, 
aspects that will be tested with further study.  The presence of resonances is 
most easily seen 
in Figure \ref{multi3} where $\mu$=10$^{-3}$.  In the range of $\delta$=4.4 to 
5.2, the points 
greatly depart from the predicted curve.  the spike at $\delta$=5.2 corresponds 
to the first and second, 
as well as the second and third planets being in 2:1 resonances.  This 
particular example shows 
that the basic dynamics of a system undergoing adiabatic mass evolution favor 
stability near 
strong resonances.  Such a process potentially could augment the current ideas 
about how resonant 
extrasolar planets such as those around GJ876 formed 
\citep{mark,phil,murray,2000ApJ...530..454R}.

\subsection{Observational Implications}

These simulations have several observational implications which can be broadly 
separated into two categories--the character and the signature of planetary 
systems around white dwarfs.  

Surviving planets that are marginally stable will suffer close approaches soon 
after 
the star evolves into a white dwarf, or possibly as early as the AGB phase.  
There are 
three possible end states for planets 
that suffer close approaches, ejection, collision, and a settling into a 
different and 
more stable configuration for all planets.  The case of two planets has been 
studied 
carefully, and for two Jovian mass planets with one planet starting at 
$\sim$5~AU, the 
probability of collision is roughly 8\%, ejection 35\%, and rearrangement 57\% 
\citep{havford}.  
We naively assume that these results hold similarly for multiple planets as 
well, since 
collisions have been shown to hold for multiple planet systems 
\citep{1997ApJ...477..781L}, 
while ejections and rearrangements should have similar probability (certainly to 
within a factor 
of 2 or so).  Ejections will leave planets that are closer to the white dwarf, 
while often 
a rearrangement will leave one or two planets with larger semi-major axes 
(up to $\sim$ 10$^3$ times greater) and one with a smaller semi-major axis 
(as close as .1 times smaller).  Collisions are potentially more exciting 
because 
as the two planets merge they essentially restart their cooling clock and as 
such will be anomalously luminous by 2 orders of magnitude 
for 10$^8$ yr \citep{burrows}.  

To estimate how many white dwarfs might have planets that collided ($F_c$) we 
can 
take the fraction of white dwarfs that have marginally stable planets and 
multiply them 
by the fraction of marginally stable planets that have collisions:
\begin{equation}
\label{frace}
F_c=f_{pl}f_{ms}f_c
\end{equation}
where $f_{pl}$ is the fraction of white dwarfs with planets, $f_{ms}$ is the 
fraction 
of marginally stable planet systems, and $f_c$ is the fraction of marginally 
stable systems 
that suffer a collision.  We can estimate the number of Jovian sized planets 
around white 
dwarfs by looking at the number of young stars that still have significant disks 
after 1 Myr, 
the approximate time to form a Jovian planet.  This has been found to be about 
50\% of 
young stars in nearby clusters \citep{2001ApJ...553L.153H}.  Several numerical 
simulations \citep[for 
example,]{2001ApJ...550..884B,1999ApJ...526..881L,quinlan,2000ApJ...530..454R,
2000ApJ...545.1044S}
point to a high frequency of marginally stable systems around stars as well as 
the 
discovery of the marginally stable planetary systems around GJ876 and HD 82943 
\citep{2001ApJ...556..296M}.  However, factors such as multiple planets with widely
different mass ratios could greatly change the effects of stability and need to be
studied further.  
We estimate this fraction to be about 50\% as well, although a large uncertainty 
is associated with 
this estimate.  Taking the results above, we estimate then that $\sim2 
(f_{ms}/0.5)$\% of young white dwarfs 
should have the product of a recent planet-planet collision in orbit. Thus we 
predict that
observations of young ($\tau \ll 10^9$ yr) white dwarfs should reveal $\sim 2$\% 
have
overluminous planet mass companions, some in orbits with semi-major axis smaller 
than
the minimum (5 AU) expected to survive the AGB phase. These planets would be 
detectable
through their significant IR excess, and should be distinguishable from brown 
dwarf companions.

A natural byproduct of the formation of Jovian planets is the existence of a 
large cloud of comets at large heliocentric distances 
\citep{1950BAN....11...91O,1999SSRv...90..301W}. 
The survival of such a cloud 
through post main-sequence evolution has been closely studied in the context of 
accounting for observed water emission in AGB stars and an explanation for 
metals 
in DA white dwarfs \citep{stern,1986ApJ...302..462A,1998ApJ...501..357P}.  The 
general result to date is 
that comets at semi-major axes greater than a few hundred AU survive the AGB 
phase. 
Then massive comets are predicted to strike the central white dwarf at a 
rate of $\sim 10^{-4} \, {\rm yr^{-1}}$, depositing fresh metals in the white 
dwarf photosphere.
Such a cometary influx can account for the DAZ phenomenon,
but has difficulty explaining some of the strongest metal line systems, because of the
short predicted residence times 
of metals in the photosphere
of these white dwarfs.
An alternate explanation for the origin of metals in 
white dwarfs is ISM accretion, where a steady drizzle of metal rich dust is
spherically accreted from the ambient ISM.
Both scenarios have difficulty explaining the frequency of DAZ white dwarfs,
and accounting for the systems with strongest metal lines and shortest residence
times, due to the high mass accretion rate required to sustain those systems.

Recent observations of the DAZ phenomenon do not seem to be consistent with 
either scenario \citep{1998ApJ...505L.143Z}.  In one DAZ, G238-44, the diffusion 
time 
for metals is 3 days, which means that neither previous scenario can explain 
the high observed metal abundances nor the stability of 
the metal lines \citep{1997ApJ...474L.127H}.  Another white 
dwarf, G29-38, has a high abundance as well as an infrared excess, possibly from 
a dust disk at small orbital radii \citep{1987Natur.330..138Z,provencal}.

The evolution of a planetary system after post-main sequence mass loss coupled 
with the presence of an Oort cloud may provide an alternative explanation for the DAZ 
phenomenon 
and in particular the peculiarities of G29-38 and G238-44.

\subsubsection{Cometary dynamics}

The mass loss during post-main-sequence is near impulsive for Oort cloud comets.
Previous work has shown that a significant fraction of any Oort cloud like
objects will survive the mass loss phase, even in the presence of mildly
asymmetric mass loss \citep{1986ApJ...302..462A,1998ApJ...501..357P}. 
The immediate result of the mass loss phase
is to leave the remaining bound objects on orbits biased towards high 
eccentricity,
but with similar initial periastrons. Orbital time scales are on the order of 
$10^6$ yr.

The number and typical size of Oort cloud objects is poorly constrained, 
canonical estimates scale to 1 km sized comets, mostly composed of low density 
ices and silicates,
with masses of $\sim 10^{16}$ gm each, with order $10^{12}$ objects per star.
Clearly there is a range of masses, and it is possible the true numbers and 
masses
of Oort cloud comets vary by several orders of magnitude from star to star. Dynamical 
effects also lead to a secular change in the amount of mass in any given Oort cloud.

External perturbations ensure a statistically steady flux of comets from the 
outer
Oort cloud into the inner system, where interactions with Jovian planets lead to 
tidal
disruption of comets (and direct collisions), scattering onto tightly bound 
orbits
restricted to the inner system, ejection from the system, and injection into 
central star
encountering orbits. For the solar system, the flux of comets into orbits 
leading
to collision with the Sun is of the order $10^{-2}$ per year,
of these a significant fraction undergo breakup before colliding with the Sun, 
with
individual fragments colliding with the Sun over many orbital periods (Kreutz 
sungrazers).
SOHO detects $\sim 10^2$ such objects per year in the solar system, or one
every 3 days or so on average. A single 1 km
comet can fragment into $\sim 10^4$ fragments with sizes of order 50 m, 
consistent
with those observed by SOHO, and consistent with the collision rates estimated 
both for the parent comets and the fragments.
Each fragment then deposits about $10^{12}$ gm into the Solar photosphere.
Note that if the typical comet were 20 km rather than 1 km, the deposition rate
would be about $10^{16}$ gm every three days.

A white dwarf has a radius about 0.01 of the solar radius. Due to gravitational 
focusing, 
the cross-section for collision for comets scattered into random orbits in the 
inner system
is linear in radius, so the collision rate expected for a white dwarf with a 
solar-like Oort cloud
is $10^{-4} \, {\rm yr^{-1}}$. However, the perturbation of the outer orbits due 
to AGB mass loss,
combined with the expansion of the outer planet orbits will drastically change 
this rate,
leading to a new, late ``heavy bombardment'' phase with significantly higher 
rates of comet
influx into the inner system. If one of the outer (Jovian mass) planets is 
scattered into
a large ($a_{fin} \gg a_{in}$) eccentric orbit after the onset 
of instability,
as we expect to happen in about 2/3 of the cases, then there will be strong 
periodic
perturbations to the outer Kuiper belt and inner Oort cloud.
About 10\% of those systems will lead to the outermost bound planet
being placed on very wide ($a_{fin} \gtorder 10^3 a_{in}$) highly eccentric
orbits, with orbital time scales comparable to the cometary 
orbital timescales. Perturbations on the Oort cloud from these planets  
lead to a persistent high flux of comets to the inner system, until the Oort 
cloud is depleted of comets.

The net effect of the dynamical rearrangement of the post-main-sequence 
planetary system is
a greatly enhanced rate of cometary influx into the inner system, 
starting $10^7$-$10^8$ years after the mass loss phase, tapering
off gradually with time on time scales of $10^8$--$10^9$ years,
leading to enhanced metal deposition to the white dwarf photosphere, and 
increased dust formation
in the inner system for some white dwarfs, depending on the final configuration 
of the outer planets.

Several processes affect the comet bombardment rate:

\begin{itemize}
\item{} a fraction of the previously stably orbiting outer Kuiper belt objects, 
that survived the AGB phase, 
are injected into the inner system by newly established dynamical resonances 
with the outer planets over
$\sim 10^8$ years;
\item{} planets ejected to the outer Oort cloud by planet-planet perturbations 
will randomise
the orbits of a small ($\sim 4(m/M)^2$) fraction of the Oort cloud comets, some 
of these
will enter the inner system providing an enhanced flux of the normal Oort comet 
infall
over $\sim 10^9$ years;
\item{} surviving inner planets, scattered to the smaller orbital radius, will 
trap the comets injected into the inner system, providing both direct tidal 
disruption
at a few AU, and providing a much higher influx of comets to very small radii
where they are tidally disrupted by the white dwarf (or in rare cases collide 
directly);
\item{} dust from tidally disrupted cometary debris will be driven to the white 
dwarf surface
by PR drag, while larger debris will be dragged in through the Yarkovsky affect, 
both on 
a timescale shorter than the WD cooling time.
\end{itemize}

We expect the Kuiper belt to be severely depleted by the post-main-sequence 
phase
\citep{stern,meinick}, however, a substantial population of
volatile depleted rocky bodies may survive the AGB phase in the outer belt.
A substantial fraction of these burnt out comets will become vulnerable to
resonant perturbations by the surviving outer planets, now in new, wider orbits.
The outer belt objects have orbital periods ($\sim 10^4$-$10^5$ years) 
that are becoming comparable to
the most shortest AGB mass loss timescales, and therefore will not generally 
expand
adiabatically in proportion to the expansion of the planetary orbits.
The solar Kuiper belt is inferred to have $\sim 10^5$ objects with size above
100 km, assuming a mass function with approximately equal mass per decade of 
mass, characteristic
of such populations,  we infer
a population of $\sim 10^{11}$ Kuiper belt objects with size of about 1 km, at 
an orbital radius of
order $10^3$ AU. Order 1\% of those will be vulnerable to the new dynamical 
resonances  
after the AGB phase, allowing for evaporative destruction and ejection, we 
estimate $\sim 10^8$
Kuiper belt objects will enter the inner system in the $10^7$-$10^8$ years after 
the AGB phase.
The rate will peak at $\sim 10^8$ years and then decline as the reservoir of 
cometary bodies
in orbits vulnerable to the new planetary resonances declines.

If there are multiple surviving Jovian planets, then the post-AGB planet-planet 
interactions
will typically leave the inner planets on eccentric orbits, leading to broader 
resonances
and a larger fraction of perturbed Kuiper belt objects.
We expect in $\sim 2/3$ of the cases where there were multiple, initially 
marginally
stable Jovian planets in the outer system, for the final configuration to have 
an 
eccentric outer planet, and a more tightly bound inner planet.

Some of the comets injected into the inner system will 
be tidally disrupted by the surviving Jovian planets, some
will be ejected, and some will be injected into the inner system and be tidally
disrupted by the white dwarf (and about 1\% of those will directly impact the
white dwarf). Dynamical time scales in the inner system are $\sim 10^2$ years,
and the probability of ejection or disruption per crossing time is of the order
of $10^{-2}$ per crossing time, assuming there is an inner planet, scattered inward
of 5 AU, matching the outer planet scattered to wider orbital radius.
So at any one time $\gtorder 10^3$ Kuiper belt objects are in the inner system. 
The rate for tidal disruption by the surviving innermost
Jovian
planets is $\sim 10^{-6} \ {\rm yr^{-1}}$ per comet. 
Tidal disruption rates due to close approaches to the white dwarf may
be as high as $\gtorder 10^{-4} \ {\rm yr^{-1}}$ per comet, the rates
are uncertain because of the possibility of non-gravitational
processes breaking up the comet and deflecting debris. With each comet massing 
about $\gtorder 10^{16}$ gm, by hypothesis, we get a flux of disrupted cometary material, from the 
Kuiper belt remnant,
assuming solar system like populations, of $10^{14}$--$10^{16}$ gm per year. This is 
volatile depleted material, by hypothesis.
Given our assumed mass function, disruption of rarer more massive comets
can sustain mass accretion rates an order of magnitude higher still for
time scales comparable to the inner system dynamical time scales, in a small
fraction of systems.

We can now compare our mechanism to the accretion rates needed to explain the 
constant, 
detected metal lines in G29-38 and G238-44, two DAZ white dwarfs with the 
highest 
measured abundances of Ca.  An accretion rate has already been quoted in the 
literature 
for G238-44, where $\sim3\times10^{17}\ {\rm g \ yr^{-1}}$ of solar abundance 
ISM would need to be accreted \citep{1997ApJ...474L.127H}.  In a volatile 
depleted case, only metals would be present 
converting to only $\sim6\times10^{15}\ {\rm g \ yr^{-1}}$ for cometary 
material.  G29-38
also has a roughly estimated value of $\sim1\times10^{19}\ {\rm g \ yr^{-1}}$ 
corresponding 
to $\sim2\times10^{17}\ {\rm g \ yr^{-1}}$ in the volatile depleted case 
\citep{provencal}.  

Both of 
these estimates were based on calculations made by \citet{d1,d2,d3}, who uses 
the ML3
 version of mixing length theory.  In fact, there are several other methods that 
can be used to model the 
convective layer of white dwarfs, including using other efficiencies of the ML 
theory and
the CGM model of convection \citep{benvenuto}.  The calculations 
differ by up to
four orders of magnitude on the mass fraction $q$ of the convection layer's base 
in white dwarfs with T$_{eff}$ similar to G29-38.  Taking values of $q$ for the 
base of the convection layer from Figures 4 
and 6 in \citet{benvenuto} and getting values for the diffusion timescale from 
tables 5 and 6 in
\citep{paquette} one can estimate what steady state accretion rate G29-38 
requires for the
different models.
The smallest rate came from ML1 theory and the largest from ML3 theory, with 
CGM having an intermediate value, giving a range of $\sim2\times10^{13}\ {\rm g 
\ yr^{-1}}$  
 to $\sim4.4\times10^{17}\ {\rm g \ yr^{-1}}$.  We favor the CGM value of
 $\sim10^{15}\ {\rm g \ yr^{-1}}$, which is consistent 
with the estimate based on observations conducted by \citet{graham}.  The rate 
for G238-44
may be more robust due to the fact that convective models converge for 
hotter white dwarfs.

Both rates are consistent with our scenario if either white dwarf has
two Jovian mass planets, one in a $\gtorder 10$ AU, {\it eccentric} orbit, and 
another
in a $\ltorder 5$ AU orbit, after rearrangement. Alternatively, their 
progenitors had
an order of magnitude richer Kuiper belt population than inferred for
the Solar system. With a post-AGB age of 
$\sim 6\times 10^8$ years,
and a mass of $\sim 0.7 \Msun$, the original main sequence star of G29-38 was 
most likely more massive than
solar and a more massive planetary and cometary system is not implausible.  
G238-44 is
almost 10$^8$ years old and would represent an object close to the 
peak of predicted cometary activity.

Our scenario may provide a consistent picture for the presence of DAZ white 
dwarfs and their anomalous properties \citep{1998ApJ...505L.143Z}. We don't expect all 
white dwarfs to have metal lines. Only about 2/3 of those which possessed marginally 
stable planetary systems containing two or more Jovians at orbital radii greater than 
$\sim 5$ AU
will be able to generate significant late cometary bombardment from the outer 
Kuiper belt
and inner Oort cloud.  Following a similar estimate as in Equation \ref{frace}, 
we predict
about 14\% of white dwarfs will be DAZs.
Further, the rate will peak after $\sim 10^8$ years, after the planet-planet perturbations
have had time to act, and then
 decline as the resevoir of perturbable comets is depeleted.  The convective 
layer of the
white dwarf will also increase by several orders of magnitude over time which 
would
create a sharp drop of high abundance DAZs with decreasing T$_{eff}$.  The drop 
would be
greatest between 12000K and 10000K where the convective layer has its steepest 
increase 
(see fig. 4 of \citet{benvenuto}).  \citet{1998ApJ...505L.143Z}, in conducting 
their survey of DAZ white
 dwarfs, estimated that $\sim$20\% of white dwarfs were DAZ and that metal 
abundance 
dropped with T$_{eff}$ sharply between 12000K and 8000K.  

We expect DAZ white dwarfs to have potentially detectable (generally) multiple 
outer
Jovian planets, whose orbits will show dynamical signatures of past 
planet-planet interaction,
namely an outer eccentric planet, and an inner planet inside the radius scoured 
clean by
the AGB phase.

\section{Discussion}
\begin{figure}
\plotone{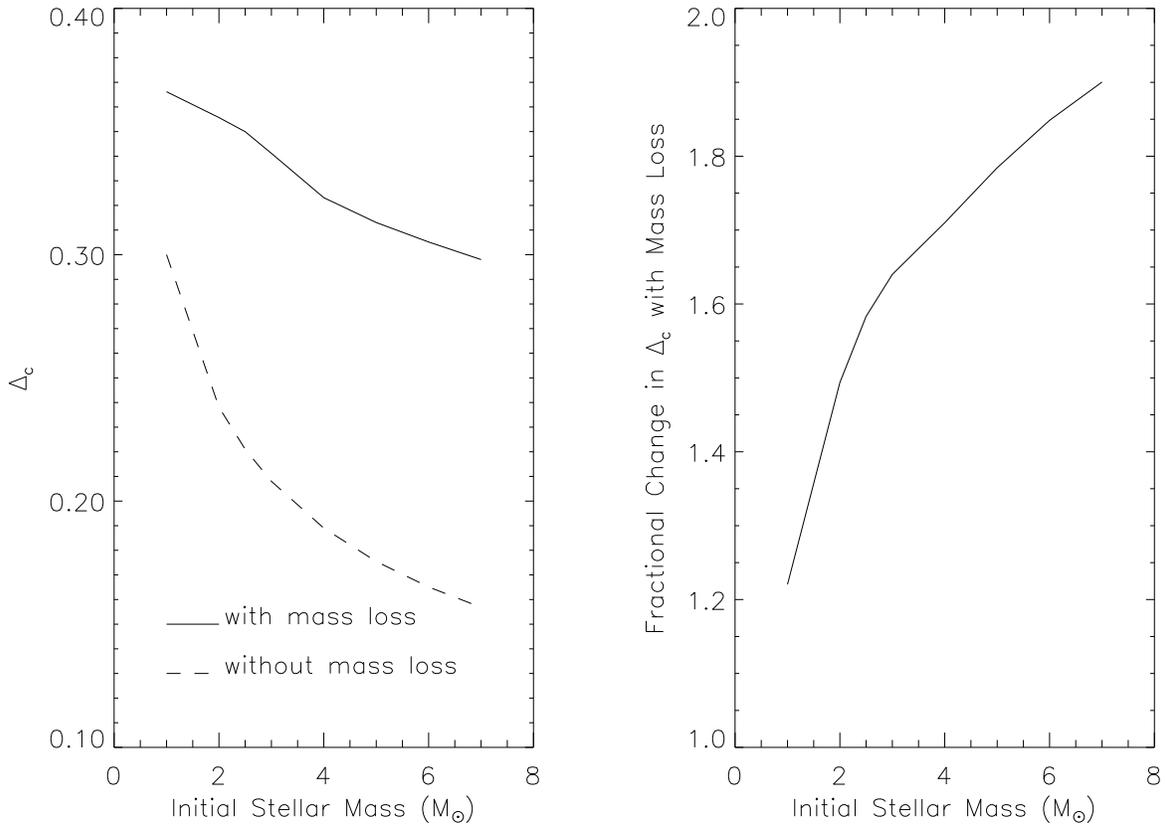}
\caption{\label{delcomp} Comparison of $\Delta_c$ with and without mass loss as 
a function of the central star's original mass.  The right figure shows the 
fractional change of $\Delta_c$ when mass loss occurs.  For both figures the 
fractional change in mass is calculated using the M$_i$-M$_f$ relation of 
\citet{2000A&A...363..647W}.}
\end{figure}

Using the above results we can compare the greatest fractional change in 
stability for two Jovian 
planets around different stars $>$ 1 M$_{\odot}$ that produce white dwarfs.  We 
took the 
initial-final mass relation of \citep{2000A&A...363..647W} and calculated 
$\Delta_c$ without 
mass loss and with mass loss ($\Delta_c^{\prime}$).  As can be seen in Figure 
\ref{delcomp}, 
the higher the initial mass star, the greater the fractional change.  This is 
expected, since 
higher mass stars lose more mass to become white dwarfs.  The best candidates 
for unstable 
planetary systems would be higher mass white dwarfs, if planet formation is 
equally efficient 
for the mass range considered here.     
The scheme conjectured in this paper, provides a method for identifying and
observing the remnant planetary systems of intermediate mass stars, which
might otherwise be hard to observe during their main sequence life time.

One can predict the change in the critical separation where two planets will 
remain 
stable based on the change in $\mu$ over time, simply by differentiating 
$\Delta_c$ with time:
\begin{equation}
\frac{d\Delta_c}{dt}=\mu^{-2/3}\frac{d\mu}{dt}.
\end{equation}
In the general case $d\mu/dt$ depends on two factors, the change in mass of the 
central 
star and the change in mass of the planets, given by 
\begin{equation}
\frac{d\mu}{dt}=\mu\left(\frac{d\ln{M_{pl}}}{dt}-\frac{d\ln{M_{\star}}}{dt}\right).
\end{equation}
For the critical separation to widen, $\mu$ must increase with time.  
Putting the two equations together gives the rate of change in $\Delta_c$:
\begin{equation}
\label{eqn1}
\frac{d\Delta_c}{dt}=\mu^{\frac{1}{3}}\left(\frac{d\ln{M_{pl}}}{dt}-\frac{d\ln{M
_{\star}}}{dt}\right).
\end{equation}

The results of our multi-planet simulations are scalable to many situations, 
but for planet systems surviving around white dwarfs we are interested in 
timescales of $\sim$10$^{10}$ yr for solar type stars to $\sim$10$^{8}$ yr 
for higher mass stars.  The highest $\delta$ we studied for $\mu$=10$^{-3}$ was 
roughly 5.2, which by Equation \ref{eqn2} corresponds to a timescale to close 
approaches 
of 10$^7$ orbits of the inner planet.  After the central star loses half of its 
mass,  
the timescale shortens to $\sim$2000 orbits.  For a planetary system with 
$\delta$=5.2 to 
be stable over the main sequence lifetime of the star, the minimum semimajor 
axis of the 
innermost planet for a higher mass star (say, 4M$_\odot$) would be 8.2~AU and 
100~AU for a 
solar-type star.  Longer integrations need to be performed to investigate the 
behavior of systems with larger values of $\delta$.    
We expect our results for time scales for onset of instability to scale to larger
$\delta$, the initial computational effort we made here limited the exploration
of slowly evolving systems with large $\delta$ in exchange for a broader exploration
of the other initial condition parameters.  It will also be instructive to model
systems with unequal mass planets, to explore the probability of ejection
and hierarchical rearrangement as a function of planetary mass ratio.

By performing numerical integrations of two planet and multiple planet systems 
we have 
shown that the stability of a system changes with mass evolution.  In the 
specific case of 
mass loss as the central star of a planetary system becomes a white dwarf, we 
have found 
that previously marginally stable orbits can become unstable fairly rapidly 
after the mass 
loss process.  Coupled with our knowledge of the survival of material exterior 
to outer planets 
such as Kuiper Belt and Oort Cloud analogues \citep{stern}, a picture of 
the evolution of circumstellar material over the latter stages of a star's 
lifetime becomes clear.

As a star reaches the RGB and AGB phases, inner planets are engulfed by both 
the expanding envelope of the star and through tidal dissipation.  The surviving 
planets 
move slowly outwards, conserving their angular momentum as the star loses its 
mass over 
several orbital periods of the planets.  The planets may sculpt the resulting 
wind of the 
giant star \citep{soker} and if they are on the very edge of stability undergo 
chaotic
 epsiodes during the AGB phase, creating some of the more exotic morphologies in the 
resulting 
planetary nebula.  When the star becomes a white dwarf, two planet 
systems that are marginally stable will become unstable and suffer close 
approaches, 
while for three or more planets the timescale to close approaches shortens by orders of 
magnitude. There are  
three possible outcomes once the planets start suffering close approaches: the 
planets collide, 
one planet is ejected, or the two planets remain but are in highly eccentric 
orbits \citep{havford}.  
One major open question is how many marginally stable systems there are, but 
there are 
indications that many, if not most, general planetary systems should be close to 
instability 
\citep{2001ApJ...550..884B,1999ApJ...526..881L,quinlan,2000ApJ...530..454R,
2000ApJ...545.1044S}   
Rocky material in the inner edge of the Kuiper Belt, which is defined by the 
last 
stable orbits with respect to the planetary system, will follow the same fate as 
marginally stable planets, suffering close approaches with the planetary system 
and becoming 
scattered into the inner system, which increases the rate of close encounters 
with the planets or 
the central white dwarf.  The surviving material at outer Kuiper Belt and Oort 
Cloud 
distances will have 
orbital periods comparable to the timescale of the central star's mass loss.  
These objects have their eccentricity pumped up by the effectively instantaneous 
change in the central star's mass, and then through interactions with planets 
create a 
new dust disk around the white dwarf, and contaminate the white dwarf 
photosphere
to an observable extent.

The sensitivity of stability to changes in mass has implications for planet 
formation as well.  
Further research on the migration of Hill stable regions while the planet/star 
mass ratio evolves
may illuminate further the general issue of how Jovian planets in the process of 
formation
become unstable to close encounters and gross changes in orbital parameters 
\citep{havford}.  
One possibility is that the mass accretion of the planets occurs at a rate fast 
enough that d$\mu/dt > 0$.  Other factors would need to be considered, in 
particular
the interplay between the onset of rapid mass accretion by the planet, and the
accretion rate from the protoplanetary disk onto the central protostar.
Gas drag and stellar mass accretion could work to stabilize orbits if the 
planets 
are embedded in a circumstellar disk, while orbital migration would change the 
relative 
separations of proto-planets.  Since the stability of multi-planet systems is 
also 
sensitive to changes in mass ratio, this could help solve problems of isolation 
for planetary 
embryos and speed up the timescale for the production of giant planet cores.

The dependence of stability on both the mass of the planet and the mass of the 
central 
star suggests that stars of different masses may be more efficient at producing 
a 
certain size planet.  This is exemplified by the fact that $\mu$ for a Jovian 
planet 
can change by an order of magnitude in either direction over the mass range of 
stars 
that might have planetary companions.  For larger mass stars, planets can be 
more tightly 
spaced and still be mutually dynamically stable, which suggests 
that when planets are forming it is easier for 
them to become dynamically isolated in disks around more massive protostars.  
For lower mass stars, there is a wider annulus in which material 
is unstable to planetary gravitational perturbations, and so 
forming planets would have a larger reservoir of material to draw from.  
Other factors that need further research would play into this result as well and 
may 
dominate over this scenario, such as a star's temperature and radiation 
pressure.  
However, such effects will tend to reinforce the conclusion that less massive 
stars should 
be more efficient at creating more massive planets while higher mass stars will 
produce more, lighter planets if they are capable of forming planets at all.  
This prediction will 
be testable as many space and ground based programs are devoting a great deal of 
effort to 
look for planetary companions to stars.  

\acknowledgements

We wish to thank Hans Zinnecker, Bill Cochran, and Brad Hansen for fruitful 
discussions.  We would also like to extend our special thanks to Dr. James 
Liebert for
helpful suggestions regarding the DAZ/planet connection.

J.D. acknowledges funding by a NASA GSRP fellowship under grant NGT5-119. S.S. 
acknowledges funding under HST grant GO-8267.

\bibliographystyle{plainnat}


\begin{thebibliography}{wd}


\bibitem[Alcock, Fristrom, \& Siegelman(1986)]{1986ApJ...302..462A} Alcock, 
C., Fristrom, C.~C., \& Siegelman, R.\ 1986, \apj, 302, 462 
\bibitem[Althaus \& Benvenuto(1998)]{benvenuto} Althaus, L.~G.~\& Benvenuto, 
O.~G.\ 1998, \mnras, 296, 206
\bibitem[Armitage \& Hansen(1999)]{1999Natur.402..633A} Armitage, P.\ J.\ 
\& Hansen, B.\ M.\ S.\ 1999, \nat, 402, 633 
\bibitem[Armitage et al.(2001)]{phil} Armitage, P.\ J. et al. 2001, preprint 
(astro-ph/0104400)
\bibitem[Barnes \& Quinn(2001)]{2001ApJ...550..884B} Barnes, R.\ \& Quinn, 
T.\ 2001, \apj, 550, 884 
\bibitem[Boss(2000)]{2000ApJ...536L.101B} Boss, A.\ P.\ 2000, \apjl, 536, 
L101
\bibitem[Butler et al.(1999)]{1999ApJ...526..916B} Butler, R.\ P. et al.\ 1999, 
\apj, 526, 916
\bibitem[Burrows et al.(1997)]{burrows} Burrows, A.\ et al.\ 
1997, \apj, 491, 856 
\bibitem[Chambers et al.(1996)]{1996Icar..119..261C} Chambers, 
J.\ E., Wetherill, G.\ W., \& Boss, A.\ P.\ 1996, Icarus, 119, 261 
\bibitem[Chu et al.(2000)]{jovwd} Chu, Y. et al.\ 2001, \apjl, 546, L61 
\bibitem[Delfosse et al.(1998)]{1998A&A...338L..67D} Delfosse, X. et al.\ 1998, 
\aap, 338, L67 
\bibitem[Duncan \& Lissauer(1997)]{1997Icar..125....1D} Duncan, M.~J.~\& 
Lissauer, J.~J.\ 1997, Icarus, 125, 1 
\bibitem[Duncan \& Lissauer(1998)]{1998Icar..134..303D} Duncan, M.\ J.\ \& 
Lissauer, J.\ J.\ 1998, Icarus, 134, 303 
\bibitem[Dupuis, Fontaine, \& Wesemael(1993)]{d1} Dupuis, J., Fontaine, G., \& 
Wesemael, F.\ 1993, \apjs, 87, 345
\bibitem[Dupuis, Fontaine, Pelletier, \& Wesemael(1993)]{d2} Dupuis, J., 
Fontaine, G., Pelletier, C., \& Wesemael, F.\ 1993, \apjs, 84, 73
\bibitem[Dupuis, Fontaine, Pelletier, \& Wesemael(1992)]{d3} Dupuis, J., 
Fontaine, G., Pelletier, C., \& Wesemael, F.\ 1992, \apjs, 82, 505
\bibitem[Ford et al.(2001)]{havford} Ford, 
E.~B., Havlickova, M., \& Rasio, F.~A.\ 2001, Icarus, 150, 303 
\bibitem[Gladman(1993)]{1993Icar..106..247G} Gladman, B.\ 1993, Icarus, 
106, 247 
\bibitem[Graham, et al.(1990)]{graham} Graham, J.~R., 
Matthews, K., Neugebauer, G., \& Soifer, B.~T.\ 1990, \apj, 357, 216
\bibitem[Griffin et al.(2000)]{2000A&AS..147..299G} 
Griffin, R.\ E.\ M., David, M., \& Verschueren, W.\ 2000, \aaps, 147, 299
\bibitem[Haisch, Lada, \& Lada(2001)]{2001ApJ...553L.153H} Haisch, K.~E., 
Lada, E.~A., \& Lada, C.~J.\ 2001, \apjl, 553, L153 
\bibitem[Hill(1886)]{hill} Hill, G.\ W.\ 1886, Acta Math., 8, 1 
\bibitem[Holberg, Barstow, \& Green(1997)]{1997ApJ...474L.127H} Holberg, 
J.~B., Barstow, M.~A., \& Green, E.~M.\ 1997, \apjl, 474, L127 
\bibitem[Kepler et al.(1990)]{1990ApJ...357..204K} Kepler, S.\ O.\ et al.\ 
1990, \apj, 357, 204 
\bibitem[Kleinman et al.(1994)]{1994ApJ...436..875K} Kleinman, S.\ J.\ et 
al.\ 1994, \apj, 436, 875
\bibitem[Koester, Provencal, \& Shipmann(1997)]{provencal} Koester, D., 
Provencal, J., \& Shipmann, H.~L.\ 1997, \aap, 320, L57
\bibitem[Kuchner et al.(1998)]{1998ApJ...508L..81K} Kuchner, 
M.\ J., Koresko, C.\ D., \& Brown, M.\ E.\ 1998, \apjl, 508, L81  
\bibitem[Laughlin \& Adams(1999)]{1999ApJ...526..881L} Laughlin, G.\ \& 
Adams, F.\ C.\ 1999, \apj, 526, 881 
\bibitem[Lin \& Ida(1997)]{1997ApJ...477..781L} Lin, D.\ N.\ C.\ \& Ida, 
S.\ 1997, \apj, 477, 781 
\bibitem[Livio \& Soker(1984)]{1984MNRAS.208..763L} Livio, M.\ \& Soker, 
N.\ 1984, \mnras, 208, 763 
\bibitem[Marcy et al.(2001)]{2001ApJ...556..296M} Marcy, G.~W., Butler, 
R.~P., Fischer, D., Vogt, S.~S., Lissauer, J.~J., \& Rivera, E.~J.\ 2001, 
\apj, 556, 296 
\bibitem[Meinick et al.(2001)]{meinick} Meinick, G.J.\, Neufeld, D.A.,\ Saavik 
Ford, K.E.,\ Hollenbach, D.J.,\ \& Ashby, M.L.N.\ 2001, Nature, 412, 160
\bibitem[Murray et al.(2001)]{murray} Murray, N.,\ Paskowitz, M.\ , \& Holman, 
M.\ 2001, preprint (astro-ph/0104475)
\bibitem[Murray \& Holman(2001)]{2001Natur.410..773M} Murray, N.~\& Holman, 
M.\ 2001, \nat, 410, 773 
\bibitem[Oort(1950)]{1950BAN....11...91O} Oort, J.~H.\ 1950, \bain, 11, 91 
\bibitem[Paquette, et al.(1986)]{paquette} Paquette, 
C., Pelletier, C., Fontaine, G., \& Michaud, G.\ 1986, \apjs, 61, 197
\bibitem[Parriott \& Alcock(1998)]{1998ApJ...501..357P} Parriott, J.~\& 
Alcock, C.\ 1998, \apj, 501, 357 
\bibitem[Pollack et al.(1996)]{1996Icar..124...62P} Pollack, J.\ B.\ et 
al.\ 1996, Icarus, 124, 62 
\bibitem[Press et al.(1992)]{numrec} Press, W.\ H.\ et al.\ 1992, Numerical 
Recipes in Fortran. (2nd Edition; New York: Cambridge University Press)
\bibitem[Quinlan(1992)]{quinlan} Quinlan, G.\ D.\ 1992, IAU 
Symp.\ 152: Chaos, Resonance, and Collective Dynamical Phenomena in the 
Solar System, 152, 25 
\bibitem[Rasio \& Ford(1996)]{1996Sci...274..954R} Rasio, F.\ A.\ \& Ford, 
E.\ B.\ 1996, Science, 274, 954 
\bibitem[Rasio et al.(1996)]{1996ApJ...470.1187R} Rasio, F.\ A.\ et al.\ 
1996, \apj, 470, 1187 
\bibitem[Rivera \& Lissauer(2000)]{2000ApJ...530..454R} Rivera, E.\ J.\ \& 
Lissauer, J.\ J.\ 2000, \apj, 530, 454 
\bibitem[Sackmann et al.(1993)]{1993ApJ...418..457S} Sackmann, I.\, 
Boothroyd, A.\ I., \& Kraemer, K.\ E.\ 1993, \apj, 418, 457 
\bibitem[Schr{\"o}der et al.(1999)]{1999A&A...349..898S} Schr{\"o}der, K.\, 
Winters, J.\ M., \& Sedlmayr, E.\ 1999, \aap, 349, 898 
\bibitem[Shu et al.(1987)]{1987ARA&A..25...23S} Shu, F.\ H., Adams, F.\ C., 
\& Lizano, S.\ 1987, \araa, 25, 23 
\bibitem[Siess \& Livio(1999a)]{1999MNRAS.304..925S} Siess, L.\ \& Livio, 
M.\ 1999, \mnras, 304, 925 
\bibitem[Siess \& Livio(1999b)]{1999MNRAS.308.1133S} Siess, L.\ \& Livio, 
M.\ 1999, \mnras, 308, 1133 
\bibitem[Snellgrove et al.(2001)]{mark} Snellgrove, M.\ D.\, Papaloizou, J.\ C.\ 
B.\ , Nelson, \& R.\ P.\ 2001, preprint (astro-ph/0104432)
\bibitem[Soker(2001)]{soker} Soker, N.\ 2001, \mnras, 324, 699
\bibitem[Soker(1999)]{1999MNRAS.306..806S} Soker, N.\ 1999, \mnras, 306, 
806 
\bibitem[Soker et al.(1984)]{1984MNRAS.210..189S} Soker, N., Livio, M., \& 
Harpaz, A.\ 1984, \mnras, 210, 189 
\bibitem[Stepinski et al.(2000)]{2000ApJ...545.1044S} Stepinski, T.\ F., 
Malhotra, R., \& Black, D.\ C.\ 2000, \apj, 545, 1044 
\bibitem[Stern et al.(1990)]{stern} Stern, S.\ A., Shull, J.\ 
M., \& Brandt, J.\ C.\ 1990, \nat, 345, 305 
\bibitem[Stoer \& Bulirsch(1980)]{bs} Stoer, J.\, Bulirsh, R.\ 1980,Introduction 
to Numerical Analysis (New York: Springer-Verlag)
\bibitem[Vassiliadis \& Wood(1993)]{1993ApJ...413..641V} Vassiliadis, E.\ 
\& Wood, P.\ R.\ 1993, \apj, 413, 641 
\bibitem[Weidemann(2000)]{2000A&A...363..647W} Weidemann, V.\ 2000, \aap, 
363, 647 
\bibitem[Weidenschilling \& Marzari(1996)]{1996Natur.384..619W} 
Weidenschilling, S.\ J.\ \& Marzari, F.\ 1996, \nat, 384, 619 
\bibitem[Weissman(1999)]{1999SSRv...90..301W} Weissman, P.~R.\ 1999, Space 
Science Reviews, 90, 30
\bibitem[Wisdom(1980)]{1980AJ.....85.1122W} Wisdom, J.\ 1980, \aj, 85, 1122 
\bibitem[Wolszczan(1994)]{1994Sci...264..538W} Wolszczan, A.\ 1994, 
Science, 264, 538  
\bibitem[Zuckerman \& Becklin(1987)]{1987Natur.330..138Z} Zuckerman, B.\ \& 
Becklin, E.\ E.\ 1987, \nat, 330, 138 
\bibitem[Zuckerman \& Reid(1998)]{1998ApJ...505L.143Z} Zuckerman, B.~\& 
Reid, I.~N.\ 1998, \apjl, 505, L143 

\begin{deluxetable}{lcccc}
\tablecolumns{4} 
\tablewidth{0pc}
\tablecaption{\label{tab1} Coefficients for Equation \ref{eqn2} derived through 
numerical simulations of three planets in circular orbits undergoing both mass 
loss(primed coefficients) and no mass loss (unprimed coefficients).  Errors 
quoted are 1 $\sigma$. The $\mu$=10$^{-7}$ case can be compared to the results 
from \citet{1996Icar..119..261C}, who determined that b=1.176~$\pm$~0.051 and 
c=-1.663~$\pm$~0.274.}

\tablehead{
\colhead{$\mu$} & \colhead{$b$}   & \colhead{$c$}    & \colhead{$b^\prime$} & 
\colhead{$c^\prime$}}
\startdata
10$^{-7}$ & 1.16 $\pm$ 0.04 & -1.6 $\pm$ 0.2 & 0.87 $\pm$ 0.05 & -1.4 $\pm$ 0.3 
\\ 
10$^{-5}$ & 1.46 $\pm$ 0.12 & -2.4 $\pm$ 0.6 & 1.14 $\pm$ 0.05 & -2.5 $\pm$ 
0.3\\
10$^{-3}$ & 2.5 $\pm$ 0.5 & -6 $\pm$ 2 & 1.2 $\pm$ 0.5 & -3 $\pm$ 2\\ \hline
\enddata
\end{deluxetable}


\end{thebibliography}

\clearpage

\end{document}